\documentclass[runningheads]{llncs}
\usepackage{amsmath}
\usepackage{multirow}
\usepackage[T1]{fontenc}
\usepackage{graphicx}
\usepackage{subcaption}
\usepackage{adjustbox} 
\usepackage{float}
\usepackage{cite}
\usepackage{booktabs}

\usepackage{color}
\usepackage{hyperref}
\hypersetup{
    colorlinks=true,      
    linkcolor=blue,       
    citecolor=blue,       
    filecolor=blue,       
    urlcolor=blue         
}
\begin{document}
\title{LIBRA: Measuring Bias of Large Language Model from a Local Context}


\author{Bo Pang\inst{1} \and
Tingrui Qiao\inst{1} \and
Caroline Walker\inst{2}\orcidID{0000-0002-9210-7651} \and
Chris Cunningham\inst{3} \and
Yun Sing Koh\inst{1}\orcidID{0000-0001-7256-4049}}
\authorrunning{B. Pang et al.}
%
\institute{School of Computer Science, University of Auckland, Auckland, New Zealand
\email{\{bpan882, tqia361\}@aucklanduni.ac.nz, y.koh@auckland.ac.nz}\\
\and
The Liggins Institute, University of Auckland, Auckland, New Zealand
\email{caroline.walker@auckland.ac.nz}\\
\and
Research Centre for Māori Health and Development, Massey University, Wellington, New Zealand\\
\email{C.W.Cunningham@massey.ac.nz}}

%
\maketitle              
\begin{abstract}
Large Language Models (LLMs) have significantly advanced natural language processing applications, yet their widespread use raises concerns regarding inherent biases that may reduce utility or harm for particular social groups. Despite the advancement in addressing LLM bias, existing research has two major limitations. First, existing LLM bias evaluation focuses on the U.S. cultural context, making it challenging to reveal stereotypical biases of LLMs toward other cultures, leading to unfair development and use of LLMs. Second, current bias evaluation often assumes models are familiar with the target social groups. When LLMs encounter words beyond their knowledge boundaries that are unfamiliar in their training data, they produce irrelevant results in the local context due to hallucinations and overconfidence, which are not necessarily indicative of inherent bias. This research addresses these limitations with a Local Integrated Bias Recognition and Assessment Framework (LIBRA) for measuring bias using datasets sourced from local corpora without crowdsourcing. Implementing this framework, we develop a dataset comprising over 360,000 test cases in the New Zealand context. Furthermore, we propose the Enhanced Idealized CAT Score (\textit{EiCAT}), integrating the \textit{iCAT} score with a beyond knowledge boundary score (\textit{bbs}) and a distribution divergence-based bias measurement to tackle the ch llenge of LLMs encountering words beyond knowledge boundaries. Our results show that the BERT family, GPT-2, and Llama-3 models seldom understand local words in different contexts. While Llama-3 exhibits larger bias, it responds better to different cultural contexts. The code and dataset are available at: \href{https://github.com/ipangbo/LIBRA}{https://github.com/ipangbo/LIBRA}.

\end{abstract}
\section{Introduction}

Large Language models (LLMs) have become a cornerstone in natural language processing (NLP) applications, providing substantial advancements in tasks ranging from chat to text generation \cite{changSurveyEvaluationLarge2024}. The use of LLMs in real-world applications has raised significant concerns about their potential biases and the impact these biases may have on different social groups \cite{gallegosBiasFairnessLarge2024}. Bias in LLMs is typically influenced by the data they are trained on, which consists of internet-sourced corpora reflecting the dominant cultural stereotypes from all over the world \cite{schickSelfDiagnosisSelfDebiasingProposal2021, navigliBiasesLargeLanguage2023}. When used in diverse cultural settings, this leads to unfair and potentially harmful outcomes. However, current research on stereotypical bias in LLMs focuses on bias based on United States' culture \cite{nangiaCrowSPairsChallengeDataset2020, parrishBBQHandbuiltBias2022, nadeemStereoSetMeasuringStereotypical2020, qianPerturbationAugmentationFairer2022}, while there is a lack of research on bias based on cultures from other parts of the world. We refer to this stereotypical bias contained in corpora in a region as the ocal bias.

\begin{figure}[t]
\begin{center}
    \includegraphics[width=0.9\textwidth]{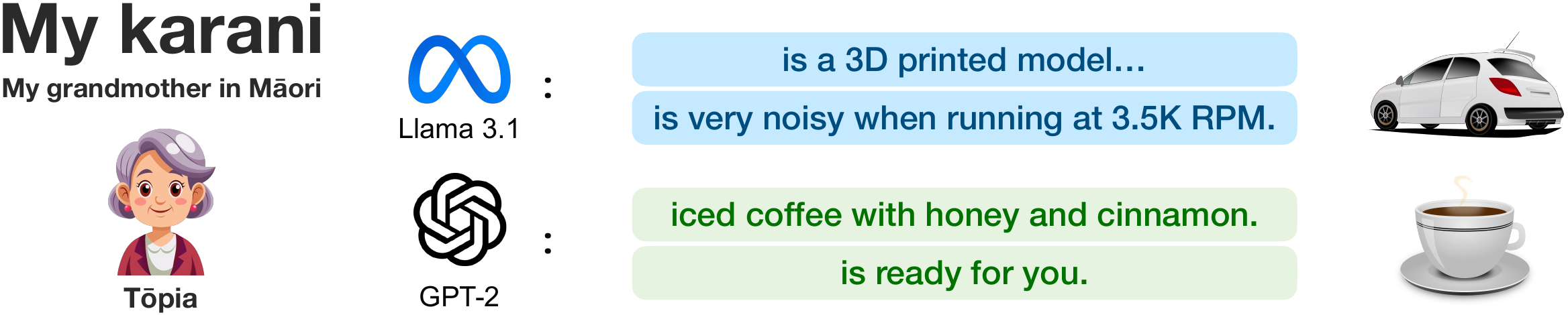}
    \caption{The comparative responses of different LLMs to prompts such as ``My karani'' or ``My karani Tōpia'', which is a transliteration between English and Māori ``My granny Tōpia'', illustrate the challenges faced in local contexts. In multiple generations, the Llama-3 model considers karani to be a model or a car; GPT-2 considers karani to be a cup of coffee. Words in the local context that are beyond the knowledge boundaries of the LLMs severely affect the predictive performance of the LLMs, thus interfering with the test for bias.} \label{karani}
\end{center}
\end{figure}

Developing methodologies to detect local biases in region-specific contexts accurately is essential, ensuring that LLMs are evaluated and improved with a sensitivity to cultural diversity. In localized contexts, a significant challenge in evaluating LLMs is the presence of words beyond the models' knowledge boundaries  \cite{renInvestigatingFactualKnowledge2023}. LLMs are often trained on extensive internet-sourced corpora, but these datasets may lack sufficient representation of region-specific terms, including those from local languages and dialects \cite{touvronLlamaOpenFoundation2023, devlinBERTPretrainingDeep2019, lanALBERTLiteBERT2020, liuRoBERTaRobustlyOptimized2019}. This deficiency can lead to unrepresentative and culturally inappropriate outputs \cite{liSurveyFairnessLarge2024}. For example, as illustrated in Fig. \ref{karani}, let LLMs generate text after ``My karani'', which contains the Māori word ``karani'' meaning grandmother. However, Llama-3 \cite{touvronLlamaOpenFoundation2023} and GPT 2 \cite{radfordLanguageModelsAre} misinterpreted ``karani'' as a model, a car, or a cup of coffee, generating unrelated content. Given these challenges, we ask the following research question: How can we effectively measure local biases in LLMs by leveraging region-specific corpora while addressing the challenges posed by local words that exceed the knowledge boundaries of LLMs?

\begin{figure}[t]
\includegraphics[width=\textwidth]{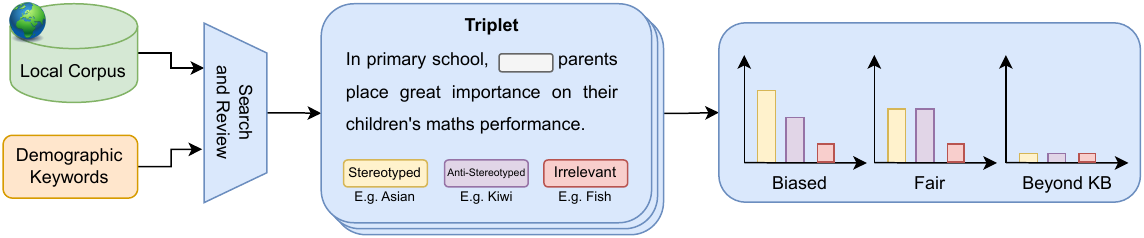}
\caption{Build and use a dataset to test the bias of Large Language Models. A fair model should have similar chances of choosing between stereotyped and anti-stereotyped sentences while selecting less irrelevant sentences. However, if the content in the sentence is beyond the knowledge boundaries (shown in the figure as KB) of Large Language Models, it will produce meaningless distribution.} \label{flowchart}
\end{figure}

To tackle the challenge of measuring the local bias of LLMs, we propose a novel framework called \textbf Local \textbf Integrated \textbf Bias \textbf Recognition and \textbf Assessment (LIBRA) for efficiently constructing datasets and measuring bias using local corpora without relying on crowdsourcing. We address the issue of words beyond LLMs' knowledge boundaries within the bias testing dataset by detecting beyond knowledge boundaries words and minimizing the effects of hallucinatory results from them during bias testing, thereby ensuring the authenticity and reliability of the test results. Additionally, we create a New Zealand-based dataset to reveal the bias of LLMs in the New Zealand context. When designing the dataset structure, we used the triplet structure of StereoSet \cite{nadeemStereoSetMeasuringStereotypical2020}, where each test case contains three similar sentences with different words describing the target social group. The description regarding target social groups requires manual annotation from crowdsourced workers. Instead of changing the descriptive words, as shown in Fig. \ref{flowchart}, we identify and replace the words that denoted the target social group in our collected corpora, thus allowing the entire dataset to be generated automatically. When dealing with sentences with words beyond the LLM's knowledge boundary, we identify local words by constructing local vocabulary lists that differ from those of general English. We add extra evaluations for local words to ensure that LLMs' interpretation of these local words matches their formal definition. A key innovation of our approach is the introduction of the Enhanced Idealised CAT Score (\textit{EiCAT}), which integrates the traditional \textit{iCAT} score \cite{nadeemStereoSetMeasuringStereotypical2020} with a beyond LLM’s knowledge boundary score (\textit{bbs}). This new metric uses the Jensen–Shannon divergence to analyze bias by examining the distribution of logits associated with stereotypical and anti-stereotypical choices in the dataset. By incorporating the \textit{bbs}, the \textit{EiCAT} score penalizes models that fail to understand local words, thus ensuring an accurate assessment of bias in localized contexts.

Our contributions are as follows.
(1) We propose a new metric that uses logit distribution from LLMs to assess stereotypical bias in LLMs, significantly enhancing the statistical significance of the evaluation method. 
(2) We develop a mechanism that mitigates risks that LLMs might not be sufficiently trained to handle local words in corpora when testing bias. 
(3)  We design a pipeline that uses local corpora to generate a dataset. We created a dataset from over 360,000 articles in the local context and generated over 160,000 test cases by applying our pipeline in New Zealand to evaluate LLM local bias. 

\section{Related Work}

\subsubsection{{Bias in Large Language Model.}}
Bias occurs when a model assumes a person possesses characteristics stereotypical of their group \cite{liSurveyFairnessLarge2024}. For instance, an LLM might use ``her'' when processing sentences that include ``nurse'', reflecting a gender bias. Such biases can lead to social injustice; for example, if a biased LLM is used for nurse CV screening, it may preferentially select females over males due to the stereotype of associating nursing with women.

\subsubsection{{Measuring Bias of Large Language Model.}}
Methods for testing LLM bias include embedding-based, generated text-based, and probability-based methods. Embedding-based methods such as WEAT \cite{caliskanSemanticsDerivedAutomatically2017} and SEAT \cite{mayMeasuringSocialBiases2019} assess similarities between vectors for target social groups and stereotype-associated vectors in text encoders' embedding. While the approach is simple, it measures upstream of LLMs and is not representative enough of downstream tasks \cite{cabelloIndependenceAssociationBias2023}. The generated text-based approach directly tests LLMs using datasets to generate potentially stereotyped results. Applicable to all LLMs, it requires only model outputs without access to text encoders or logits. However, using classifiers to analyze these outputs can introduce their own biases \cite{pozzobonChallengesUsingBlackBox2023}. Probability-based methods, exemplified by StereoSet \cite{nadeemStereoSetMeasuringStereotypical2020} and CrowS-Pairs \cite{nangiaCrowSPairsChallengeDataset2020}, use pseudo-log likelihoods to assess the probability of word generation, ideally suited for open-source LLMs or APIs providing logits data. The pseudo-log likelihoods used by StereoSet \cite{nadeemStereoSetMeasuringStereotypical2020} are represented as follows: $L(S)=\frac{1}{|M|}\sum_{t \in M}\log P(t|U;\theta)$\label{mlmfs}, where $S$ is a sentence, $M$ and $U$ are masked and other words in $S$, and $\theta$ is other parameters in the model. The Stereotype Score (\textit{ss}) measures the proportion of biased sentences preferred by the model. In contrast, the Language Model Score (\textit{lms}) reflects the selection percentage of non-irrelevant terms, showing the model's language ability. StereoSet \cite{nadeemStereoSetMeasuringStereotypical2020} integrates these into the Idealized CAT Score (\textit{iCAT}): $iCAT = lms \cdot \frac{min((100 - ss), ss)}{50}$\label{icat}. Our proposed metric improves bias measurement by analyzing the distributional distance between logits for stereotypical and anti-stereotypical responses, unlike traditional metrics focusing solely on the highest probability choice. Our approach detects model preferences certainty, providing a more comprehensive understanding of model bias across various behaviours.

Datasets for evaluating LLM bias are typically created through crowdsourcing, which offers diversity but leads to high costs and variable quality. Alternative methods like filling sentence templates with varying words produce monotonous content and lack syntactic diversity \cite{parrishBBQHandbuiltBias2022, gallegosBiasFairnessLarge2024}. Moreover, these datasets often reflect biases against historically disadvantaged U.S. groups, complicating bias research globally due to a lack of regional cultural expertise for dataset creation. Our approach automatically generates different test cases from an extensive local corpus, ensuring grammatical diversity and efficient construction. Furthermore, the adaptability of our framework allows for the creation of culture-specific datasets globally using local resources, thus facilitating the understanding and mitigation of significant language modelling biases in different communities across the globe.

\section{LIBRA Framework for Measuring Local Bias}
\label{methodology}

The measurement of local bias in LLMs necessitates a local dataset paired with a metric. This section first outlines the dataset's structure and application for comprehensive bias assessment in LLMs. We then introduce our novel metric, \textit{EiCAT}, and describe a pipeline for dataset creation and bias evaluation. Finally, we address the challenge of words beyond LLMs' knowledge boundaries in local contexts and propose solutions to mitigate their impact on bias assessment.

\subsubsection{Dataset Construction.}

To address problems with existing datasets that test for bias in LLMs, we propose a methodology for creating bias detection datasets in LLMs without crowdsourcing, using local corpora. We define the whole local corpora as $\mathcal{O}$. It consists of articles denoted as $\mathcal{A} = [A_1,\ldots,A_n]$. We predefine a set of seeding keywords $\mathcal{K}$, using target social groups $\mathcal{G}$ to guide keyword selection. For each social group $G \in \mathcal{G}$, we identify directly associated seed keywords $(K, G) \in \mathcal{K}$.  These keywords are then expanded by finding terms in a text encoder's embedding space that show high similarity, adding to $\mathcal{K}$. To expand the keywords further, we use association rule learning \cite{borgeltImplementationFPgrowthAlgorithm2005} for each $(K, G) \in \mathcal{K}$ to discover words that often appear together with keyword $K$ and add to $\mathcal{K}$ because words often appear with keywords are often potential keywords. The keyword set $\mathcal{K}$ is available for searching potentially biased sentences.

Directly searching the corpora $\mathcal{O}$ using all keywords from $\mathcal{K}$ is not sensitive enough because each keyword $K$ is linked to a specific target social group $G$. To refine this, we employ fuzzy clustering \cite{mcinnesHdbscanHierarchicalDensity2017} to group articles with similar topics into clusters $(C, G) \in \mathcal{C}$, where each cluster is associated with a target social group. We then use a targeted keyword set $\mathcal{K}_{G_i} = \{K|(K, G) \in \mathcal{K} \text{ and } G=G_i\}$ to search sentences in corresponding articles $\mathcal{A}_{G_i} = \{A | A \in \mathcal{A} \text{ and } A \in C \text{ and } (C, G) \in \mathcal{C} \text{ and } G=G_i\}$. We extract all sentences $S$ in $\mathcal{A}_{G_i}$ that contain the keywords in $\mathcal{K}_{G_i}$ to form the set of potentially biased sentences $\mathcal{S}$. The sentences in $\mathcal{S}$ screened by local cultural experts can be used to construct a dataset for testing the bias of LLMs.

We structure our dataset as a series of triplets, each comprising an original stereotyped sentence from the corpora ($S$), an anti-stereotyped sentence ($S_{p}$), and an unrelated sentence ($S_{u}$) for ensuring LLMs not erroneously favouring any particular response due to content irrelevance. Each sentence in triplet $T \in \mathcal{D} = [T_1,\ldots, T_n]$ is tokenized, with $S = [t_1,\ldots,t_n]$. We identify target social group tokens $\omega = [t_j, t_{j+1},\ldots,t_k]$, and for $S_{p}$ and $S_{u}$, we retain unmodified tokens $U = [t_1,\ldots, t_{j-1}, t_{k+1},\ldots, t_n]$. The sentences are then formed by replacing $\omega$ with demographic terms for $S_{p}$ and unrelated terms for $S_{u}$, respectively, resulting in $S = U_{\text{left}} + \omega + U_{\text{right}}$, $S_{p} = U_{\text{left}} + M_p + U_{\text{right}}$, and $S_{u} = U_{\text{left}} + M_u + U_{\text{right}}$.

\subsubsection{Use Dataset to Measure LLMs Bias.}
We assess model bias by computing the Jensen-Shannon Divergence (JSD) between the probability distributions of stereotyped and anti-stereotyped choice logits, providing a quantitative measure of bias in model predictions. For instance, consider two models processing the same test case: one assigns a probability of 99\% to the stereotyped option and 1\% to the anti-stereotyped option, while another assigns 51\% and 49\%, respectively. Traditional assessment methods might label both models as biased. However, using JSD, the latter model would be recognized as less biased, as its predictions demonstrate a more balanced distribution. Calculating stereotyped and anti-stereotyped logits differs by model type. Masked Language Models (MLMs) like BERT \cite{devlinBERTPretrainingDeep2019}, RoBERTa \cite{liuRoBERTaRobustlyOptimized2019}, and ALBERT \cite{lanALBERTLiteBERT2020} predict tokens based on their left and right context. In contrast, Causal Language Models (CLMs), such as GPT \cite{radfordLanguageModelsAre} and Llama \cite{touvronLlamaOpenFoundation2023}, generate the next token sequentially \cite{gallegosBiasFairnessLarge2024}. We use $L(\cdot)$ to indicate the likelihood of LLMs generating specific responses within contexts. For MLMs, this is calculated as the pseudo-log-likelihood of logits (Equation \ref{mlmfs}). For CLMs, $L(\cdot)$ is defined as the sum of logits, excluding the left part of unmodified tokens, which do not influence sentence comparison:
\begin{equation}
    \label{clmfs}
    L(S) = \frac{1}{|S-U_{\text{left}}|} \sum_{i = |U_{\text{left}}| + 1}^{|S|} \log P(S[i]|S[0],\ldots, S[i - 1];\theta),
\end{equation}
where $S[i]$ is the $i$-th token in the sentence $S$, $\theta$ is other parameters in the model. 

For each test case, \textit{i.e.}, a triplet, $L(S)$ quantifies the model's tendency across three sentences in triplets. After evaluating all triplets, we compile the $L(S)$ values for stereotyped sentences into distribution $\mathcal D_s$ and anti-stereotyped sentences into $\mathcal D_a$. We then measure bias between these distributions using the Jensen-Shannon Divergence, $\text{JSD}(\mathcal D_a||\mathcal D_s)$. A $\text{JSD}(\mathcal D_a||\mathcal D_s)$ closer to 0 indicates less bias by showing similarity between the distributions, and vice versa.

\subsubsection{Dataset Creation Pipeline.}
We provide a pipeline to generate a dataset from organized corpora articles, enabling local context bias measurement in LLMs.

\noindent\textit{{Keywords Augment.}} We use a broad set of seed keywords related to target social groups to search for sentences in our dataset. These keywords, crucial within their categories, are expanded using the LLM2Vec tool \cite{behnamghaderLLM2VecLargeLanguage2024}, which allows decoder-only LLMs to generate embedding spaces and find similar words within the corpora. Further expansion is achieved through association rule learning \cite{borgeltImplementationFPgrowthAlgorithm2005}, identifying frequently co-occurring terms from the corpora as \textit{Augmented} keywords.

\textit{Clustering and Allocating Social Groups.}
Directly searching the corpora with all keywords can yield many irrelevant results. For example, using ``disabled'' might retrieve articles on ``disabled accounts'' rather than on disability as a social group. To enhance relevance, we transform the text into vectors using LLM2Vec ~\cite{behnamghaderLLM2VecLargeLanguage2024} and apply dimensionality reduction with UMAP \cite{mcinnesUMAPUniformManifold2018}. We then categorize the corpora into topics using HDBSCAN \cite{mcinnesHdbscanHierarchicalDensity2017}, assigning texts to the nearest clusters through Iterative Clustering \cite{liYouHearPeople2023} if they do not fit an existing category. Each topic is linked to relevant social groups using two steps: summarizing cluster contents with an incremental summary technique \cite{changBOOOOKSCORESYSTEMATICEXPLORATION2024} to manage context length and prompting LLMs to identify associated social groups.

\textit{{Search for Sentences.}}
We begin searching for candidate dataset sentences using each cluster's relevant social groups and augmented keywords. We divide cluster texts into individual sentences and then search these using keywords associated with the cluster's target social groups. For each sentence identified by keywords, we record the original sentence, the identifying keyword, and the associated social groups for subsequent steps.

\textit{{Compile Dataset.}}
Candidate dataset sentences are refined through further processing. We use a perturbating strategy to construct anti-stereotyped and irrelevant options by modifying demographic noun keywords. Perturbating involves choosing antonyms for keywords with clear opposites (e.g., ``man'' vs ``woman''). For keywords without direct opposites, such as ``European'', we select terms from the same social group keyword set $\mathcal{K}_{G_i}$ that are distant in the text encoder's embedding space. Local culture experts review all triplets to ensure an accurate representation of stereotypes and anti-stereotypes.

\subsubsection{Effect of Beyond Knowledge Boundary Words in Local Context.}
\label{beyondkb}

Local context-specific words can disrupt bias testing. This section outlines a method to identify and exclude words beyond an LLM's knowledge boundary from test sentences $\mathcal{S}$. If LLMs fail to comprehend certain local terms, those test cases are marked invalid to prevent meaningless tests. Additionally, we factor the proportion of unrecognized local words into our quality metric for the model.

For each test case sentence set $\mathcal{S}$, we compile a vocabulary list of all appearing words $\mathcal{V}$. We then exclude words and their variants found in English dictionaries, denoted as $\mathcal{V}_w$, resulting in $\mathcal{V}' = \mathcal{V} \setminus \mathcal{V}_w$. The $\mathcal{V}'$ represents words beyond LLMs' knowledge boundaries within $\mathcal{S}$.
To verify if LLMs understand a word $W \in \mathcal{V}'$, we compare the model's interpretation with its formal definition, avoiding direct prompts that could induce model hallucinations and false positives in the results \cite{renInvestigatingFactualKnowledge2023}. For each word $W \in \mathcal{V}'$, we use prompt template $P_1$ to extract its definition $D_1$ from an LLM and compare it with the official definition $D_2$. If $D_1$ and $D_2$ align, the test case is valid; otherwise, any triplets with $W$ are marked invalid. The verification template $P_2$ can be applied on any LLM since this task is independent of the tested model, allowing using more capable models if necessary.

For each word $W \in \mathcal{V}'$, with $S_w$ being the sentences containing $W$, we define a binary function $f(W, S_w)$ to assess if a language model accurately understands $W$ in $S_w$; $f(W, S_w) = 1$ means $D_1$ and $D_2$ matches, and $0$ indicates not matching. To refine LLM bias evaluation in localized contexts, we introduce Beyond LLM's Knowledge Boundary Score (\textit{bbs}):

\begin{equation}
    bbs = \frac{\sum_{W\in \mathcal{V}' }f(w,S_w)}{|\mathcal{V}'|}.
\end{equation}
The Enhanced Idealized CAT Score (\textit{EiCAT}), which incorporates measures of bias, language model capacity and knowledge boundaries, is as follows:
\begin{equation}
\begin{aligned}
EiCAT = lms \cdot [\alpha \cdot (1 - \text{JSD}(\mathcal D_A||\mathcal D_S)) + (1 - \alpha) \cdot bbs],
\end{aligned}
\end{equation}
where $\alpha$  is a weighting parameter that adjusts the contribution of \textit{JSD} and \textit{bbs} to the overall \textit{EiCAT} score, allowing customization on the bias measurement.

The \textit{EiCAT} score ranges from 0 to 1 and quantitatively evaluates fairness and effectiveness in local contexts of LLMs, with higher scores indicating better performance and less bias. We set the weighting parameter $\alpha$ equal to the \textit{bbs} value to dynamically balance the influence between bias assessment and the model's understanding of local-specific vocabulary. When \textit{bbs} is high, indicating the model effectively comprehends local terms, $\alpha$ increases, emphasising the bias measurement component more. Conversely, if \textit{bbs} is low, suggesting a limited understanding of local words, $\alpha$ decreases, increasing the relative weight of the \textit{bbs}. This approach ensures the \textit{EiCAT} score appropriately penalizes models for bias and lack of contextual understanding, providing a more accurate and fair assessment of their capabilities in localized contexts.

\section{New Zealand Context Dataset Construction}
\label{nzdataset}

New Zealand's bicultural foundation, comprising indigenous Māori and European settlers, is further diversified by subsequent immigration. Māori and Pacific Peoples have historically faced poor socio-economic outcomes stemming from colonization and cultural marginalization \cite{sibleyEthnicGroupStereotypes2011}. While the Treaty of Waitangi settlements have addressed some historical grievances, the residue of biases persists \cite{yogarajanTacklingBiasPretrained2023}.
\begin{figure}[h]
\begin{center}
    \begin{subfigure}[t]{0.42\textwidth}
        \centering
        \includegraphics[width=\textwidth]{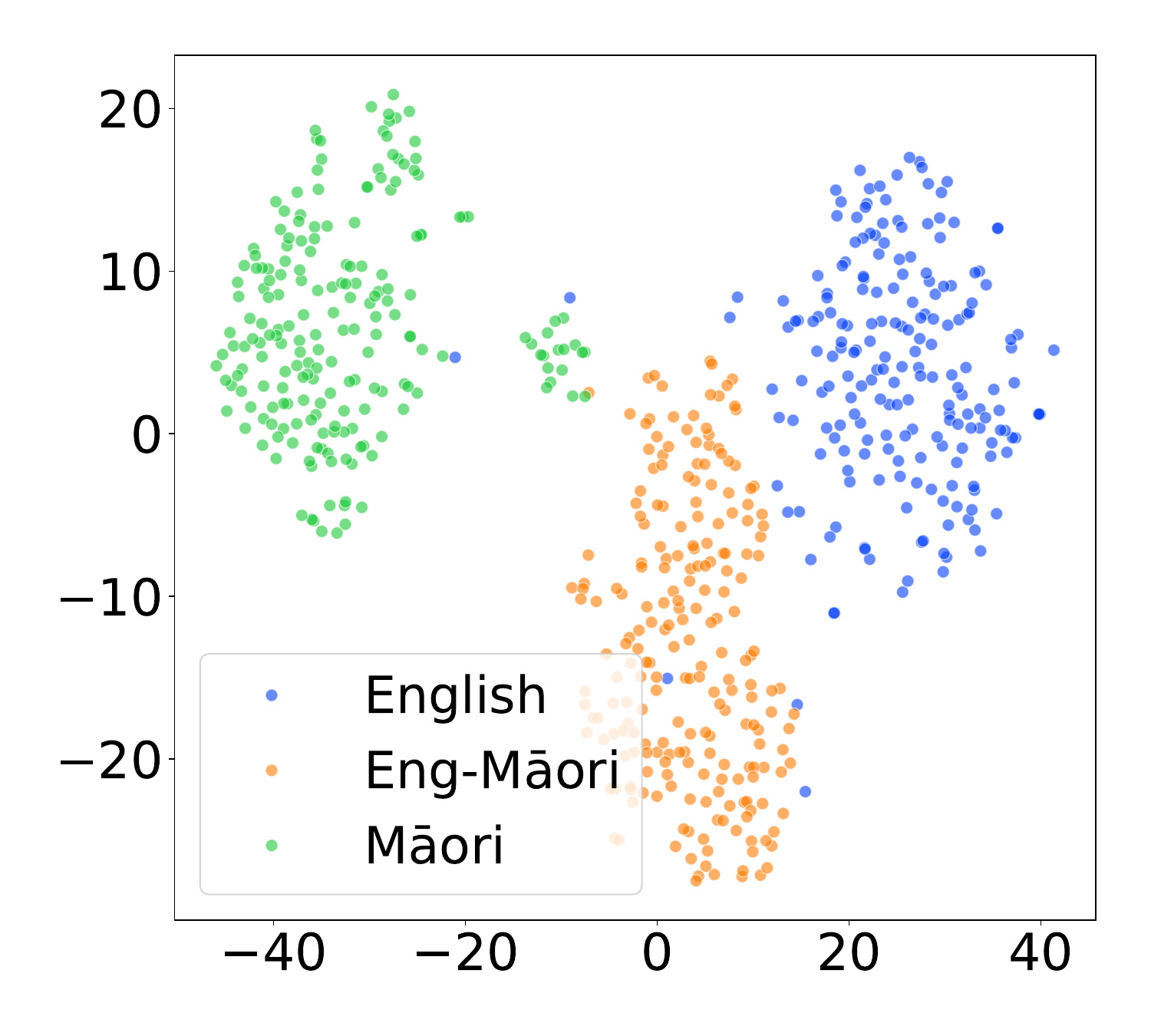}
        \caption{}
        \label{fig:a}
    \end{subfigure}
    \begin{subfigure}[t]{0.45\textwidth}
        \centering
        \includegraphics[width=\textwidth]{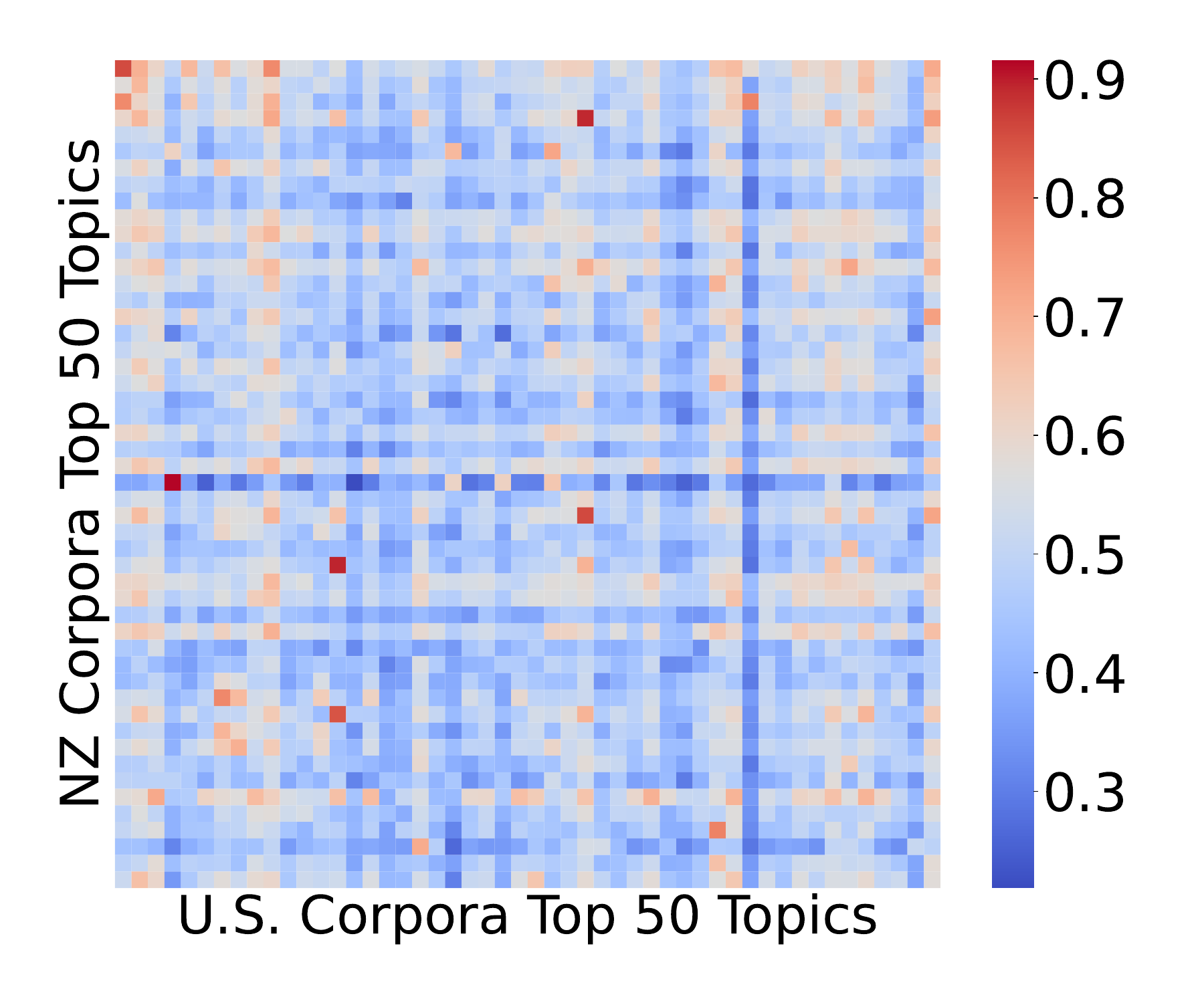}
        \caption{}
        \label{fig:b}
    \end{subfigure}
    \caption{Visualisations of Contextual Diversity. Fig. (a) shows BERT embeddings clustering purely English, mixed English-Māori, and solely Māori sentences separately, indicating LLMs' distinct treatment of linguistic variations that could limit effective output across them. Fig. (b) reveals a heatmap of similarities among the top 50 topics from U.S. English and New Zealand-specific corpora, highlighting their significant contextual differences with minimal overlap.}
    \label{visualisations}
\end{center}
\end{figure}

We apply our pipeline in a New Zealand context to explore local bias and LLM knowledge boundaries, as Fig. \ref{visualisations} illustrates the differences between New Zealand and general English contexts. To enhance syntactic diversity, we include both text-source and oral-source texts in our corpora \cite{chanDynamicDevelopmentSpeaking2015}. For privacy, we perform Named Entity Recognition (NER) \cite{liSurveyDeepLearning2022}. Depending on the source, the raw corpora may require filtering out bias-irrelevant content such as weather forecasts, jokes, and puzzles. Finally, we use 367,384 news articles and broadcast transcripts collected from New Zealand local media to analyze language use associated with various social groups in New Zealand. Our dataset, comprising 167,712 sentences and triplets, assesses bias across eight target groups as summarized by \cite{meadeEmpiricalSurveyEffectiveness2022}. 
It addresses the significant presence of Te Reo Māori ``borrow-words'' in the NZ English corpora and whose limited presence in LLM training data often poses challenges.

The distribution of sentences across different social groups in our dataset mirrors their prevalence in local corpora. Age-related content is predominant, making up 73.43\% of the total, reflecting its frequent discussion and relevance as a social factor. Gender and race/ethnicity also feature prominently, accounting for 10.78\% and 10.65\%, respectively, highlighting their importance in the discourse. Lesser represented groups include sexual orientation (2.94\%), physical appearance (0.96\%), disability (0.78\%), nationality (0.02\%), and religion (0.45\%), indicating these topics are less frequent or more context-specific in the corpora.

Our dataset recognises ``noise'' as a test case that does not precisely target the intended social groups. Instead of individual assessments, we analyze the collective distribution of these cases using horizontal comparisons across models. This method ensures that our comparative analysis remains effective even in noisy scenarios where models consistently select the same stereotyped option for fluency. The consistent response to poorly flowing sentences confirms that our metric accurately captures model biases, offering a reliable framework for evaluating biases across various cultural contexts.

\section{Results And Discussions}

Here, we assess which words in our dataset are first beyond the knowledge boundaries of several prominent open-source LLMs, including the BERT family, GPT-2, and Llama 3 family (Version 3.1). Next, we present the results of applying our New Zealand context dataset and evaluation metrics to these models. All experiments use the Transformers library \cite{wolf2019huggingface} and are performed on an NVidia A100 with 80GB of video memory.

\subsubsection{LLMs Knowledge Boundaries in New Zealand Context.}
Our experiments involved analyzing words from the New Zealand corpus beyond the knowledge boundaries of tested models, revealing significant performance variations. Evaluating the $bbs$ in Table \ref{tab:res} shows all tested LLMs achieve low scores, indicating poor comprehension of non-English words in the New Zealand context. The Llama family model outperformed others, while GPT-2 struggled with non-English words from our dataset, and the BERT family models' performance was moderate, positioned between Llama and GPT-2.

\begin{table}[htbp]
\centering
\caption{Performance Metrics of Different LLMs on the New Zealand Context Dataset. This table presents evaluation metrics scaled to a 0-100 range for readability, covering Language Model Score (\textit{lms}), Jensen-Shannon Divergence (\textit{JSD}), Beyond Knowledge Boundary Score (\textit{bbs}), traditional Idealized CAT Score (\textit{iCAT}) using the StereoSet dataset for comparison, and Enhanced Idealized CAT Score (\textit{EiCAT}). The models are categorized into Theoretical Language Models (TLMs), Masked Language Models (MLMs), and Causal Language Models (CLMs). We also assess the largest model variants in specific contexts: New Zealand young people (OV) using the Our Voices dataset and the Malaysia context (Malay) to understand bias within these unique cultural settings. Theoretical models are included to illustrate extreme cases for reference. The best performance is highlighted in bold for each metric and category.}
\label{tab:res}
\begin{tabular}{llrrrrr}
\toprule
 \makebox[0.05\textwidth][l] & \makebox[0.2\textwidth][l]{\textbf{Model}} & \makebox[0.12\textwidth][r]{\textbf{\textit{lms}}}& \makebox[0.12\textwidth][r]{\textbf{\textit{JSD}}}& \makebox[0.12\textwidth][r]{\textbf{\textit{bbs}}}&\makebox[0.12\textwidth][r]{\textbf{\textit{iCAT}}}& \makebox[0.12\textwidth][r]{\textbf{\textit{EiCAT}}}\\ \midrule
\multirow{4}{*}{\rotatebox{90}{\parbox{1.5cm}{\centering \textbf{TLMs}}}}
  & RandomLM                 &  66.67&   0.00&   0.00&   66.67&   0.00 \\ 
  & IdealLM                  & \textbf{100.00}&   0.00&   0.00& \textbf{100.00}&   0.00 \\
  & LocalIdealLM             & \textbf{100.00}&   0.00& \textbf{100.00}& \textbf{100.00}& \textbf{100.00} \\
  & StereotypedLM            & \textbf{100.00}& \textbf{100.00}& \textbf{100.00}&   0.00&   0.00 \\ \midrule
\multirow{8}{*}{\rotatebox{90}{\parbox{1.5cm}{\centering \textbf{MLMs}}}}
  & BERT-base       &  96.04&  45.44&   3.96&  71.60&  5.73  \\
  & BERT-large      &  96.73&  47.27&   \textbf{4.11}&  70.28&  \textbf{5.91}  \\ 
  & RoBERTA-base   &  96.89&  39.11&   1.67&  \textbf{84.43}&  2.58  \\
  & RoBERTA-large      &  \textbf{97.46}&  39.83&   2.28&  81.58&  3.51  \\ 
  & ALBERT-base-v2    &  89.11&  27.56&   0.08&  44.92&  0.12  \\
  & ALBERT-large-v2   &  88.26&  22.49&   1.90&  39.70&  2.94  \\
  & ALBERT-xlarge-v2  &  86.23&  \textbf{21.37}&   1.22&  41.88&  1.87  \\
  & ALBERT-xxlarge-v2 &  91.28&  35.58&   0.46&  44.44&  0.69  \\ \midrule
\multirow{5}{*}{\rotatebox{90}{\parbox{1.5cm}{\centering \textbf{CLMs}}}}
  & GPT-2-base        &  57.82&   \textbf{0.49}&   1.07&  37.80&  1.23  \\
  & GPT-2-medium     &  57.30&   0.73&   1.75&  51.85&  1.98  \\
  & GPT-2-large      &  59.67&   1.48&   1.37&  59.83&  1.61  \\
  & GPT-2-xl         &  58.78&   1.37&   1.45&  \textbf{60.68}&  1.68  \\ 
  & Llama-3-8b         &  \textbf{77.48}&   3.35&   \textbf{7.31}&  59.16& \textbf{10.72}  \\ \midrule
\multirow{5}{*}{\rotatebox{90}{\parbox{1.5cm}{\centering \textbf{OV}}}}
  & BERT-large      &  \textbf{96.45}&  40.86&   0.00&      70.28&   0.00         \\ 
  & RoBERTA-large      &  95.96&  32.93&   \textbf{9.52}&      \textbf{81.58}&  \textbf{14.39}      \\ 
  & ALBERT-xxlarge-v2 &  94.63&  28.59&   4.76&      44.44&   7.51    \\ 
  & GPT-2-xl         &  63.49&   \textbf{1.92}&   0.00&      60.68&   0.00     \\ 
  & Llama-3-8b         &  80.38&   2.86&   4.76&      59.16&   7.36      \\ \midrule
\multirow{5}{*}{\rotatebox{90}{\parbox{1.5cm}{\centering \textbf{Malay}}}}
  & BERT-large      & \textbf{100.00}&  63.77&  \textbf{17.86}&      70.28&  \textbf{21.14}      \\ 
  & RoBERTA-large      &  99.22&  39.39&  10.71&      \textbf{81.58}&  15.93        \\ 
  & ALBERT-xxlarge-v2 &  92.97&  39.90&   7.14&      44.44&  10.15    \\ 
  & GPT-2-xl         &  68.81&   \textbf{1.72}&   0.00&      60.68&   0.00       \\ 
  & Llama-3-8b         &  83.93&   2.49&   7.14&      59.16&  11.41\\ 
\bottomrule
\end{tabular}
\end{table}

\subsubsection{LLM Bias in New Zealand Context.}

Analysis of the New Zealand dataset indicates that BERT and RoBERTa outperform ALBERT in language model scores, showing stronger text completion abilities. However, their elevated \textit{JSD} and \textit{bbs} scores suggest they exhibit more bias due to greater differences in logit distribution between stereotyped and anti-stereotyped responses. This comparison highlights the inherent biases of BERT and RoBERTa relative to ALBERT. Notably, the \textit{JSD} values cannot be directly compared across masked and causal models. GPT-2 displays lower biases but weaker language capabilities. In contrast, the Llama-3 model achieves higher \textit{EiCAT} scores by effectively handling local words, demonstrating the best performance of all causal language models in the New Zealand context.

\begin{figure}[t]
\centering
\begin{subfigure}[b]{0.3\textwidth}
    \includegraphics[width=\textwidth]{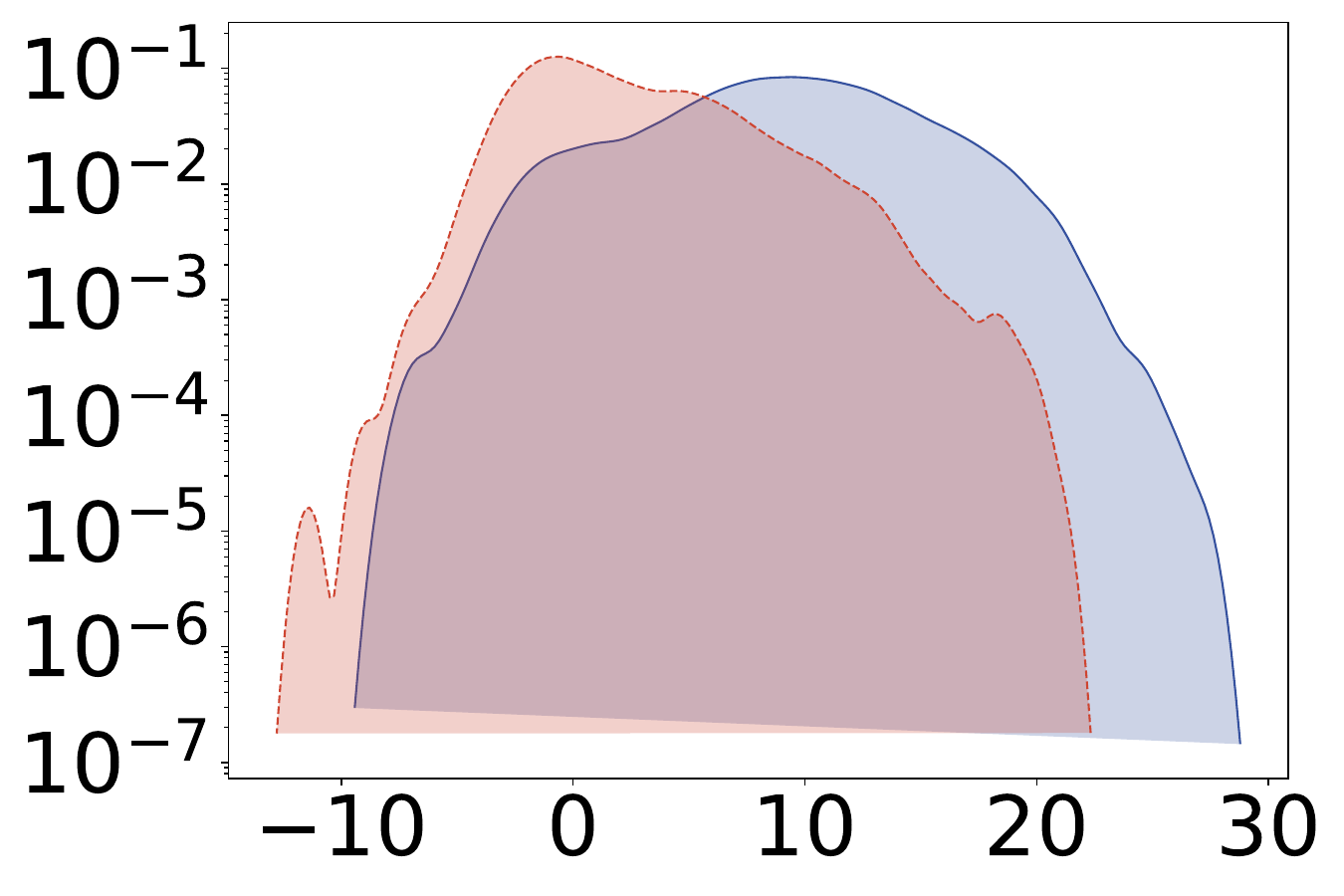}
    \caption{BERT-large}
    \label{fig:sub1}
\end{subfigure}
\hfill 
\begin{subfigure}[b]{0.3\textwidth}
    \includegraphics[width=\textwidth]{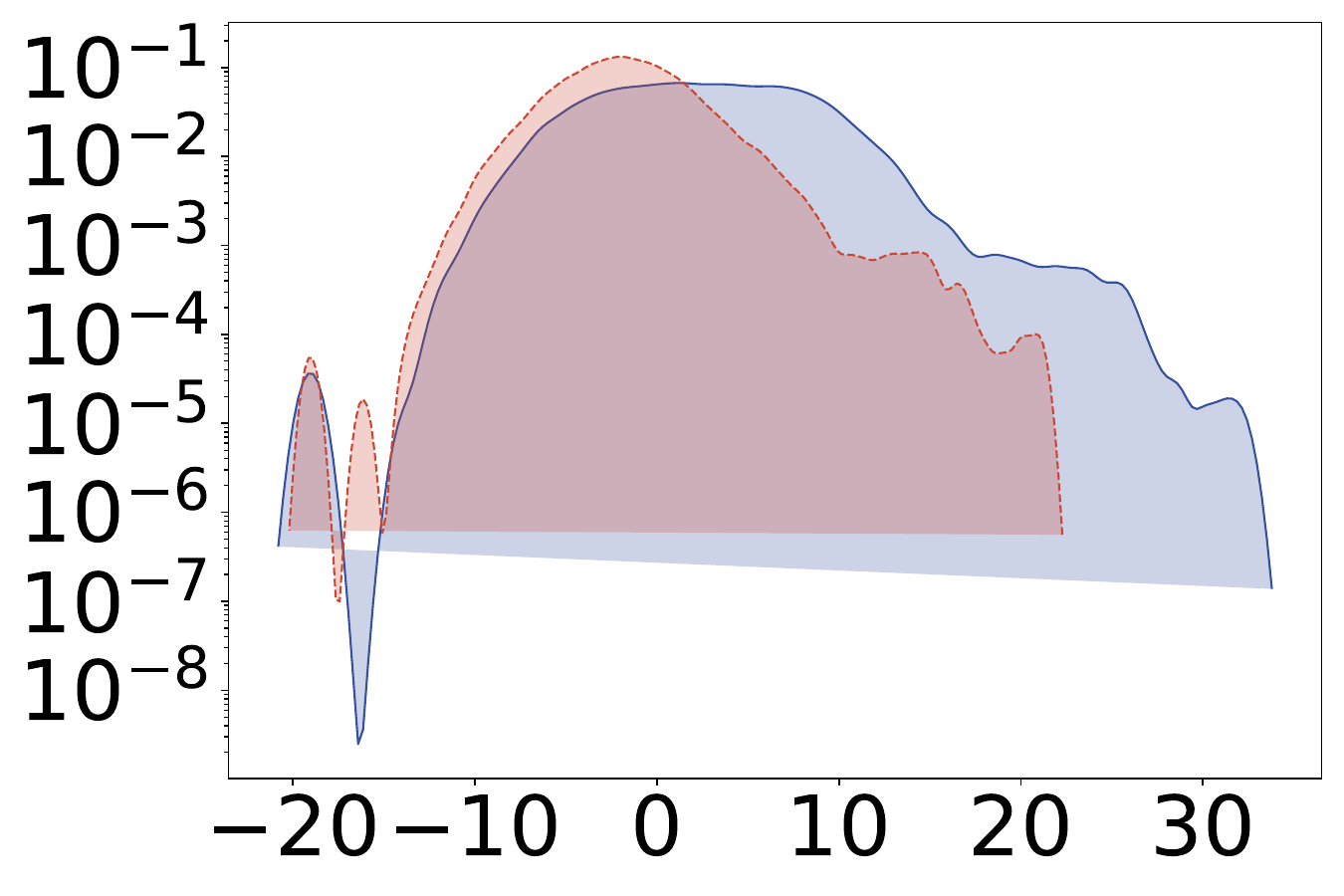}
    \caption{ALBERT-xxlarge}
    \label{fig:sub2}
\end{subfigure}
\hfill
\begin{subfigure}[b]{0.3\textwidth}
    \includegraphics[width=\textwidth]{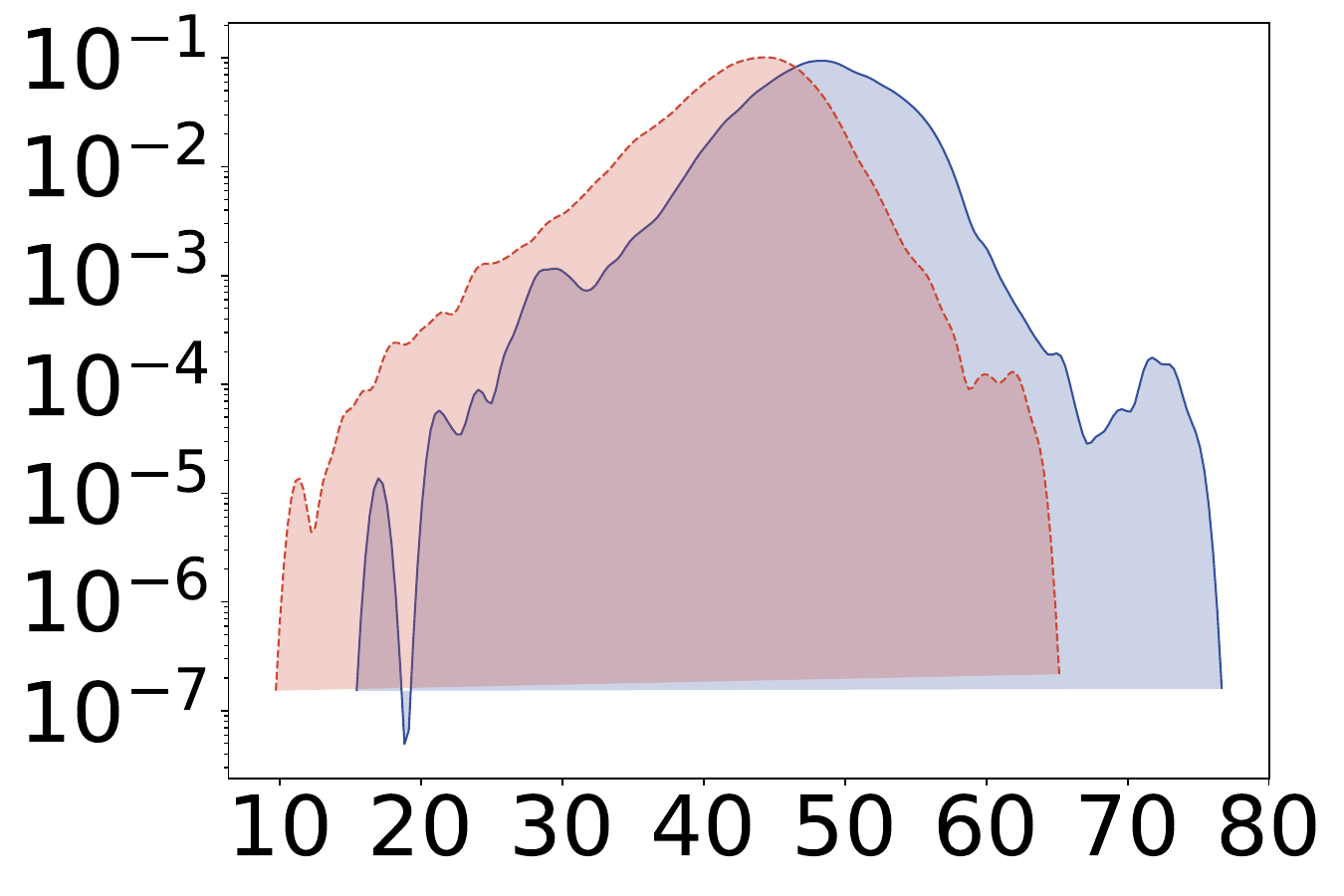}
    \caption{RoBERTA-large}
    \label{fig:sub3}
\end{subfigure}


\begin{subfigure}[b]{0.3\textwidth}
    \includegraphics[width=\textwidth]{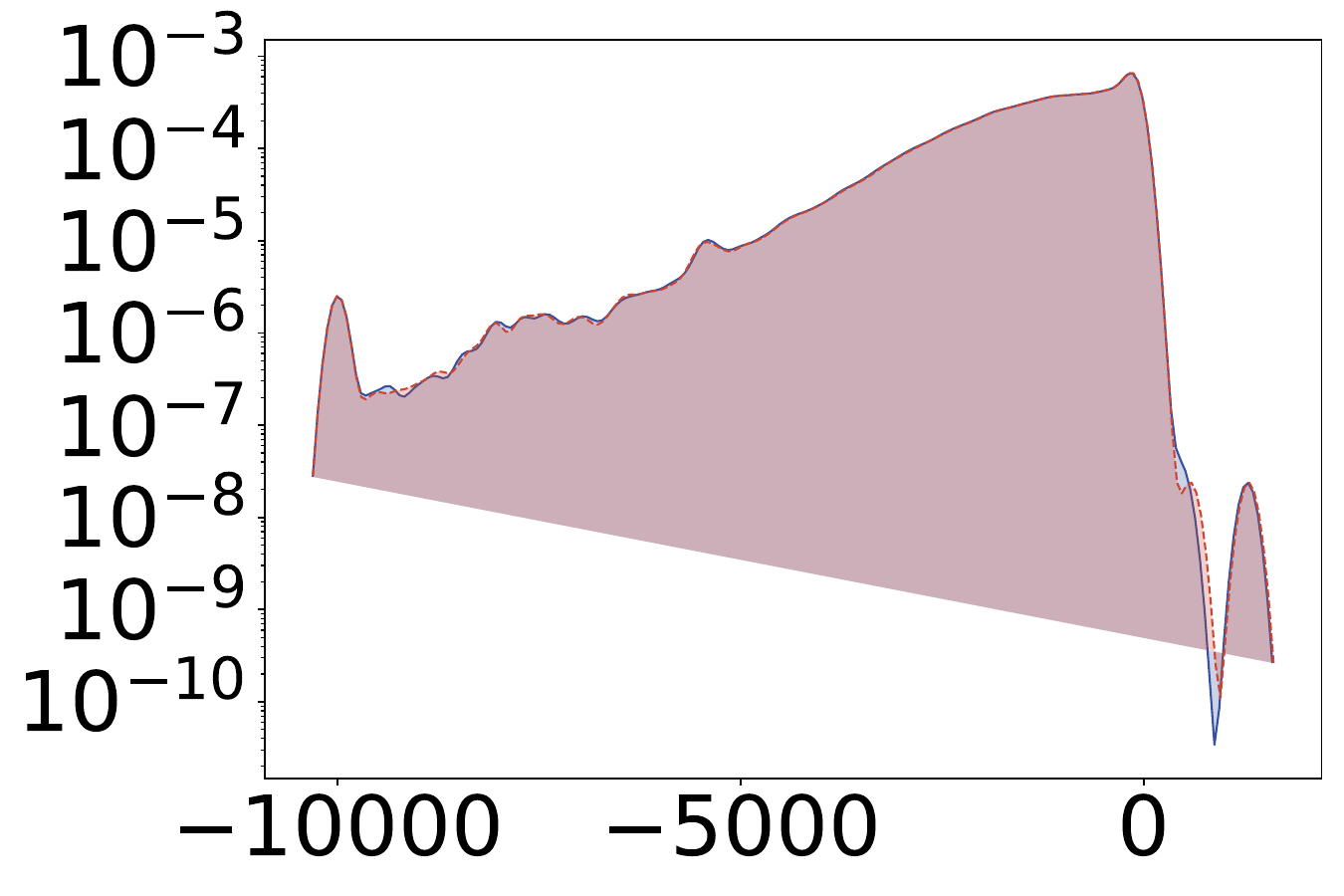}
    \caption{GPT-2-xl}
    \label{fig:sub4}
\end{subfigure}
\hfill
\begin{subfigure}[b]{0.3\textwidth}
    \includegraphics[width=\textwidth]{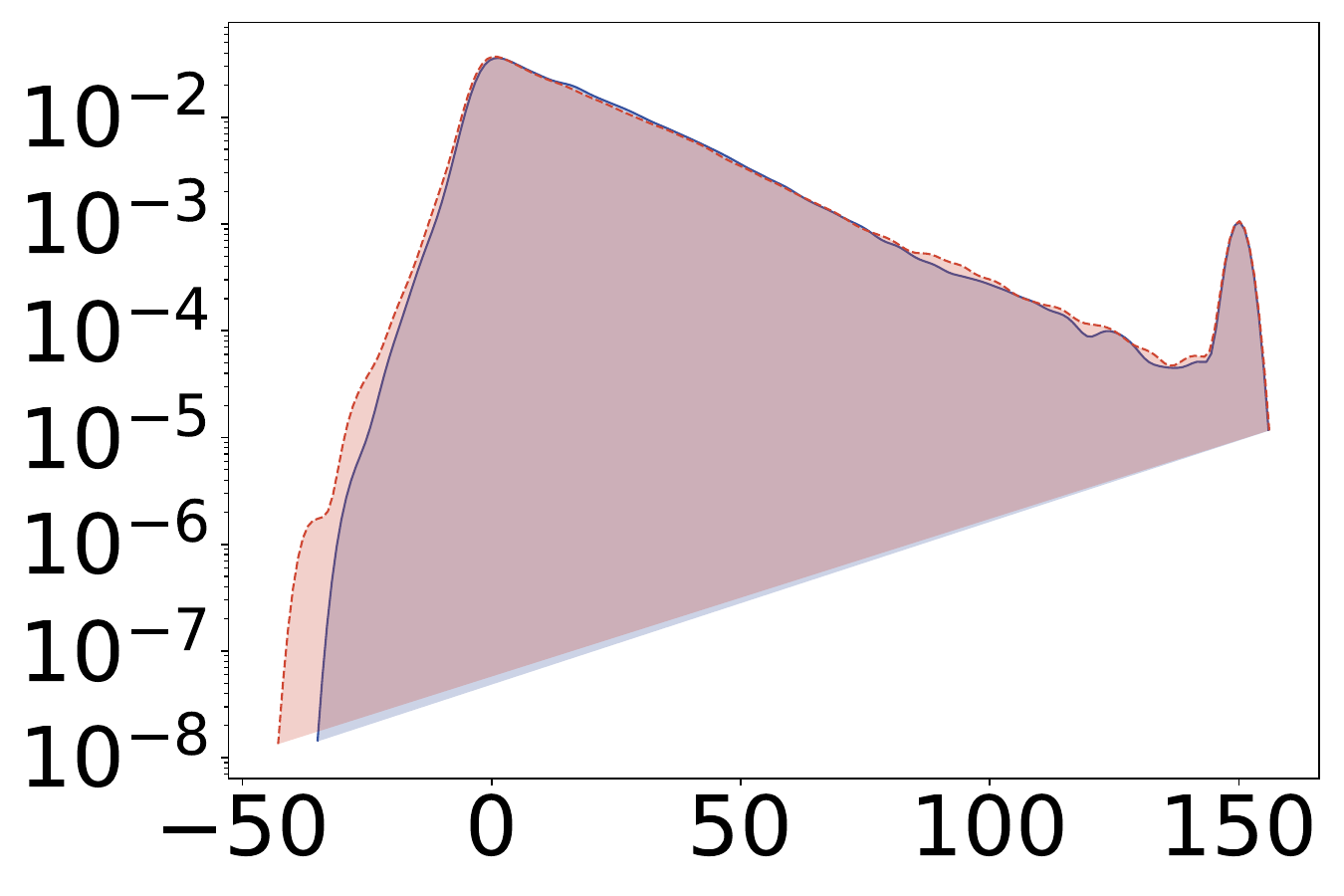}
    \caption{Llama-3-8B}
    \label{fig:sub5}
\end{subfigure}
\hfill
\begin{subfigure}[b]{0.3\textwidth}
    \includegraphics[width=\textwidth]{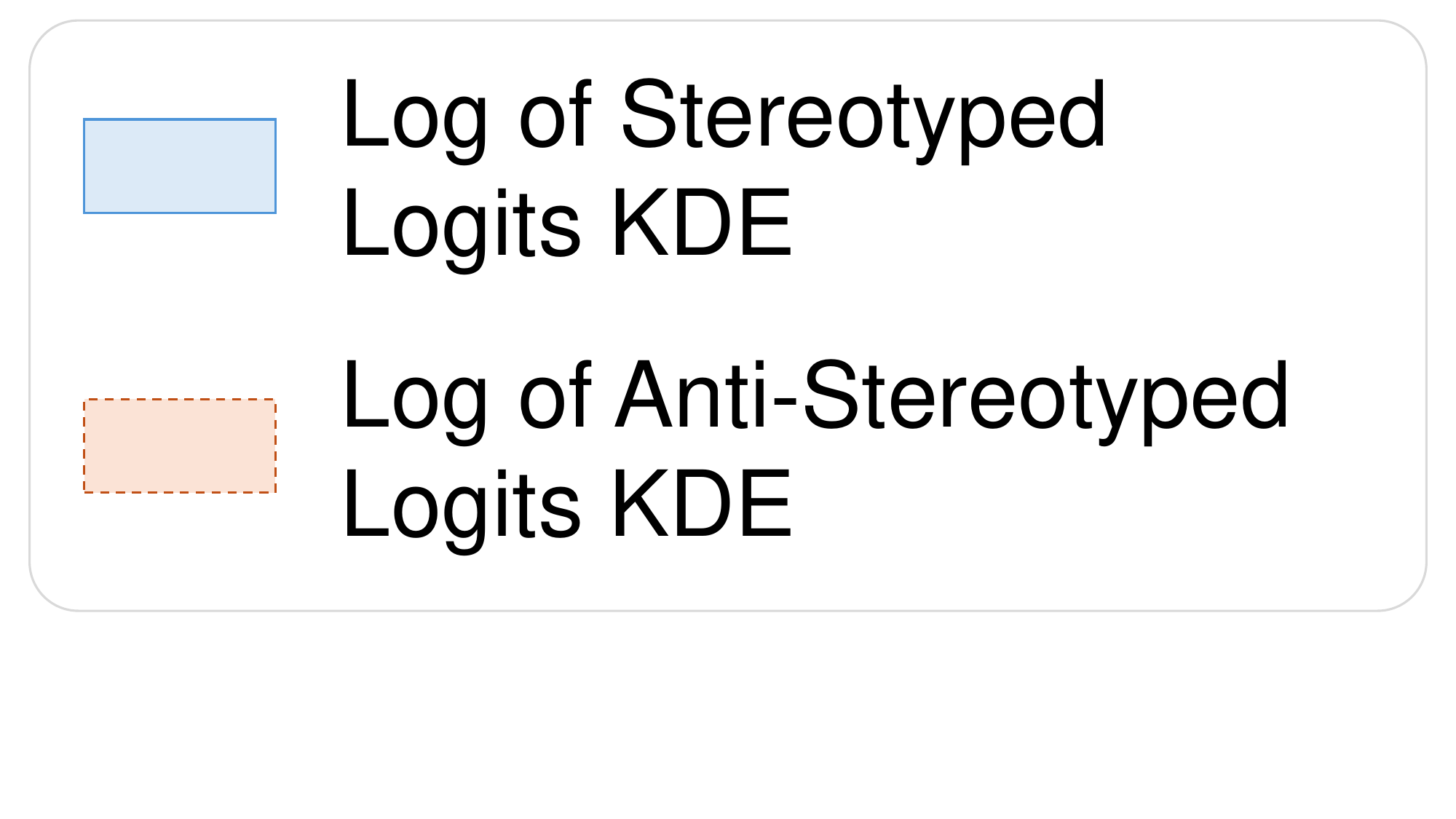}
    \label{fig:sub6}
\end{subfigure}
\caption{Comparison of kernel density estimation (KDE) plots for the log-transformed density of logits across the largest size of tested LLMs in the New Zealand Context. Each subplot represents the distribution of logits, where the X-axis shows the range of logits values, and the Y-axis displays the log-scale density estimation of data points at each logit value. Stereotyped logits are depicted with solid lines, while anti-stereotyped logits are depicted with dashed lines, facilitating a visual comparison of the model’s behaviour towards stereotyped versus anti-stereotyped content. A larger divergence represents the model with a larger bias.}
\label{fig:mainfigure}
\end{figure}

\subsubsection{LLM Bias in Other Contexts.}

We evaluate LLM biases using two culturally distinct contexts to validate our framework's adaptability. The first, \textit{Our Voices} dataset, collects natural language data from young New Zealanders, reflecting contemporary sociolinguistic trends among youth \cite{OurVoicesProjects}. The second context comes from the MEN dataset \cite{chanthranMalaysianEnglishNews}, consisting of 200 Malaysian local news. Malaysian English is influenced by Malay, Chinese, and Tamil, featuring unique lexical and syntactic variations \cite{chanthranMalaysianEnglishNews}. These datasets enable a comprehensive analysis of LLM performance across varied linguistic environments.

We assess the largest model sizes in these contexts. BERT shows the highest linguistic competence but the highest bias. RoBERTa consistently attains high \textit{EiCAT} scores and robust \textit{bbs}, indicating its effective handling of local vocabulary. ALBERT's performance is slightly inferior, with reduced linguistic capabilities and less bias. Llama-3 outperforms other causal models, achieving the highest \textit{EiCAT} scores. Remarkably, some models score zero due to a \textit{bbs} of zero, indicating a failure to recognize local-specific vocabulary.

\subsubsection{Discussion}
Our framework provides a refined measure of models' ability to handle culturally specific content without bias. By integrating the bias assessment with the $bbs$, $EiCAT$ captures the fairness of the model's predictions without hallucination and overconfidence and its capacity to understand and process culturally unique vocabulary. This dual focus is crucial for ensuring cultural fairness in LLMs, particularly for underrepresented languages. Furthermore, the framework empowers researchers globally to efficiently evaluate LLM biases in specific local contexts, guiding the broader use of LLMs globally.

Our experimental results highlight the varying performance of different models within different contexts. The higher $EiCAT$ scores of the Llama-3 model indicate that it better understands words in the local context and reduces bias, representing it as the most suitable model to handle the task in the New Zealand context. Even though Llama-3 did not achieve the best scores in all contexts, it is still the strongest overall model among the causal language models. This performance is due to their advanced architectures and more diverse training datasets. However, even the best-performing models like Llama are not entirely free from biases or limitations in handling words beyond their knowledge boundaries, underscoring the need for continuous improvements in model training.

BERT, RoBERTa and Llama-3, despite achieving high $lms$, display a greater degree of bias. The exact opposite is the case for ALBERT and GPT-2. As with StereoSet, we find that models with higher \textit{lms} have higher \textit{JSD}, which indicates that more linguistically competent models tend to be more biased. The relationship is because LLMs objectively have biases, and the more linguistically competent the model, the more accurately it can represent the biases inherited from the training data. This also suggests we accurately reflect the bias by comparing logit distribution distances.

\section{Conclusion}

We proposed the LIBRA framework to measure biases in large language models (LLMs) within local contexts. By leveraging diverse and expansive corpora, we developed a New Zealand context dataset, relying on robust methods to gather and curate extensive local data rather than traditional crowdsourcing approaches. The Enhanced Idealized CAT Score (\textit{EiCAT}), introduced within the framework, integrates traditional bias metrics with a beyond knowledge boundary score (\textit{bbs}) and a distributional divergence-based assessment, offering a comprehensive evaluation tool. Applying this framework to the New Zealand context, we utilized data sourced from New Zealand media to create a dataset comprising over 160,000 test cases. Our analysis revealed that while models such as Llama-3 exhibit certain biases, they demonstrate a stronger capacity to handle culturally specific vocabulary and terminology. By testing knowledge boundaries, we also highlighted the importance of formal definitions and structured data in assessing LLMs' capabilities with underrepresented languages such as Māori and Malay. These findings underscore the necessity of incorporating localized and contextually rich data in evaluating and improving LLM fairness. LIBRA offers a scalable, adaptable approach for researchers globally, promoting the development of more fair and culturally sensitive language technologies.

\subsubsection{Acknowledgement.} The authors thank the Growing Up in New Zealand rangatahi / youth who participated in the Our Voices study, conducted by the Our Voices study team. The Our Voices study was designed by Susan M.B. Morton (GUiNZ Foundation Director and Principal Investigator of Our Voices 2019-2022) and the Our Voices study team. The authors acknowledge the contributions of the original Our Voices team and study investigators: Susan M.B. Morton, Rizwan Asghar, Polly E. Atatoa Carr, Chris Cunningham, Daniel Exeter, Sarah Knowles, Yun Sing Koh, Christopher Nixon, Elizabeth R. Peterson, Avinesh Pillai, Chris Schilling, Caroline Walker, Joerg Wicker, Kane Meissel (Principal Investigator 2023-2025), and those who have contributed at various points throughout the study. The views reported in this paper are those of the authors and do not necessarily represent the views of the Our Voices Investigators or Growing Up in New Zealand. Financial support: The Our Voices study was funded by an Endeavour grant (UOAX1912) by the Ministry of Business, Innovation and Employment (2019-2025). This funding also provided a travel grant awarded to the first author of this manuscript. Funders had no role in the design, analysis or writing of this article. Conflict of interest: None. Ethics of human subject participation: This study was conducted according to the guidelines laid down in the Declaration of Helsinki and all procedures involving human subjects were approved by the Ministry of Health’s Northern B Health and Disability Ethics Committee. Consent was obtained from all participants and their parent or guardian.

\bibliographystyle{splncs04}
\bibliography{ecir}

\end{document}